\documentclass[a4paper,fleqn]{cas-dc}

\usepackage[numbers]{natbib}

\def\tsc#1{\csdef{#1}{\textsc{\lowercase{#1}}\xspace}}
\tsc{WGM}
\tsc{QE}
\tsc{EP}
\tsc{PMS}
\tsc{BEC}
\tsc{DE}

\begin{document}
	\let\WriteBookmarks\relax
	\def\floatpagepagefraction{1}
	\def\textpagefraction{.001}
	\shorttitle{Stripe-like vortex flow}
	\shortauthors{S. V. Filatov et~al.}
	
	\title [mode = title]{Generation of stripe-like vortex flow by noncollinear waves on the water surface}
             	
	\author[1,2]{S. V. Filatov} []
	\ead{fillsv@issp.ac.ru}
	\author[1,2]{A. V. Poplevin}[]
    \cormark[1]
	\ead{faraldos@issp.ac.ru}
	\author[1,2]{A. A. Levchenko}[]
	\ead{levch@issp.ac.ru}
	\author[2,3]{V. M. Parfenyev}[]
	\ead{parfenius@gmail.com}
	
	\address[1]{Institute of Solid State Physics, Russian Academy of Sciences, 2 Academician Ossipyan str., 142432 Chernogolovka, Russia}
	\address[2]{Landau Institute for Theoretical Physics, Russian Academy of Sciences, 1-A Akademika Semenova av., 142432 Chernogolovka, Russia}
	\address[3]{National Research University Higher School of Economics, Faculty of Physics, Myasnitskaya 20, 101000 Moscow, Russia}
	
	\cortext[cor1]{Corresponding author}

	\begin{abstract}
		We have studied experimentally the generation of vortex flow by gravity waves with a frequency of $2.34$ Hz excited on the water surface at an angle $2 \theta = \arctan(3/4) \approx 36^\circ$ to each other. The resulting horizontal surface flow has a stripe-like spatial structure. The width of the stripes $L=\pi/(2k \sin \theta)$ is determined by the wave vector $k$ of the surface waves and the angle between them, and the length of the stripes is limited by the system size. It was found that the vertical vorticity $\Omega$ of the current on the fluid surface is proportional to the product of wave amplitudes, but its value is much higher than the value corresponding to the Stokes drift and it continues to grow with time even after the wave motion reaches a stationary regime. We demonstrate that the measured dependence $\Omega(t)$ can be described within the recently developed model that takes into account the Eulerian contribution to the generated vortex flow and the effect of surface contamination. This model contains a free parameter that describes the elastic properties of the contaminated surface, and we also show that the found value of this parameter is in reasonable agreement with the measured decay rate of surface waves.
	\end{abstract}
		

	\maketitle
	
	\section{Introduction}
	
The horizontal mass transport induced in a fluid by surface waves is a long-standing problem of both fundamental and practical interest. In laboratory experiments, this transport is studied by observing the motion of passive particles advected by the flow. The corresponding Lagrangian velocity V (averaged over fast wave oscillations) for a single plane wave was calculated by Longuet-Higgins up to the second-order in wave amplitude in Ref.~\cite{longuet1953mass}. Later, his results were generalized to the case of several surface waves propagating in arbitrary directions \cite{nicolas2003three}. The qualitative difference is that in the case of noncollinear waves, the Lagrangian velocity has nonzero vertical vorticity $\Omega=\partial_x V_y - \partial_y V_x$, which makes it possible to produce horizontal drift currents having various spatial structures \cite{filatov2015generation,filatov2016nonlinear,abella2020spatio}.

The vertical vorticity of the induced current on the fluid surface can be described by the sum of two terms \cite{parfenyev2018influence, parfenyev2019formation}
\begin{equation}\label{eq:1}
	\Omega = \Omega_S + \Omega_E.
\end{equation}
The first term corresponds to the generalization of the Stokes drift in ideal fluid \cite{stokes1847theory} to the case of noncollinear waves. It is the result of nonlinear Lagrangian dynamics during one time period of oscillations and it
does not produce any contribution to the mean velocity of fluid in the Eulerian description \cite{longuet1969nonlinear}. The second term, which we call the Eulerian contribution, on the contrary, is the mean velocity of fluid. It is excited by a force, which is localized in the narrow viscous sublayer near the fluid surface and is produced due to hydrodynamic nonlinearity  (it is also known as the virtual wave stress \cite{longuet1969nonlinear, weber2001virtual}).

These two contributions have different steady-state amplitudes and characteristic time-scales. The situation was analyzed in detail theoretically in Ref.~\cite{parfenyev2020large}, and for two surface waves propagating at an angle $2 \theta$ to each other on the surface of infinitely deep water, it was found that the ratio of steady-state amplitudes is equal to $\Omega_E/\Omega_S = 1/ \sin \theta$. The Stokes drift contribution instantly tracks the changes in the wave motion and its characteristic time $T_S$ can be estimated as half the decay time of the waves since the Stokes drift is quadratic in the wave amplitudes. The Eulerian contribution is generated by the surface stress and then it spreads downwards due to viscous diffusion. Therefore, one can estimate its characteristic time as $T_E = L^2/(\pi^2 \nu)$, where $\nu$ is the fluid kinematic viscosity, $\pi/L = 2 k \sin \theta$, $L$ is the horizontal size of the slow current produced due to the second-order nonlinearity and $k$ is the wave number of the excited surface waves.

For typical experimental conditions $T_S \ll T_E$ due to additional dissipation of the wave motion near the system boundaries \cite{parfenyev2019formation, landau1987course} and surface contamination \cite{parfenyev2018influence, campagne2018impact}. The latter factor also significantly affects the amplitude of the Euler contribution $\Omega_E$ at the fluid surface \cite{parfenyev2018influence, parfenyev2019formation}, but in all cases, the theory predicts its universal dependence on time
\begin{equation}\label{eq:scaling}
	\Omega_E (t) \propto \mathrm{erf} \left[ \sqrt{t/T_E}\right], \quad \mathrm{erf}[x]= \dfrac{2}{\sqrt{\pi}} \int_{0}^{x} d \xi e^{-\xi^2}.
\end{equation}
This expression remains valid in the case of a liquid of finite depth $d \gg 1/k$, but now its applicability is limited in time, $t \ll d^2/\nu$. At these times, the Eulerian current does not have time to penetrate to the bottom due to viscous diffusion and therefore does not feel its presence.

The presented description is valid only if the Eulerian currents are sufficiently weak and their nonlinear interactions can be neglected. This means that the effective Reynolds number $Re = \Omega_E(t) T_E$ for slow currents should remain small. The goal of the presented experimental study is to test theoretical predictions. We change the amplitudes of the surface waves and show that the relation (\ref{eq:scaling}) holds for small amplitudes and it breaks down when the degree of nonlinearity increases ($Re \geq 10$). In comparison with the previous work \cite{parfenyev2019formation}, we managed to increase the ratio $T_E/T_S$ by an order of magnitude, which significantly increased the accuracy. Also, we show that the amplitude of vorticity at the surface and its spatial structure can be explained using the model presented in Ref.~\cite{parfenyev2020large} if we take into account the possible contamination of the fluid surface. This model contains a free parameter that describes the elastic properties of the contaminated surface. This parameter also affects the decay rate of surface waves and we demonstrate that the found value is in reasonable agreement with the performed measurements.
	
	\section{Experimental Methods}
	
\begin{figure}
		\centerline{\includegraphics[width=0.9\linewidth]{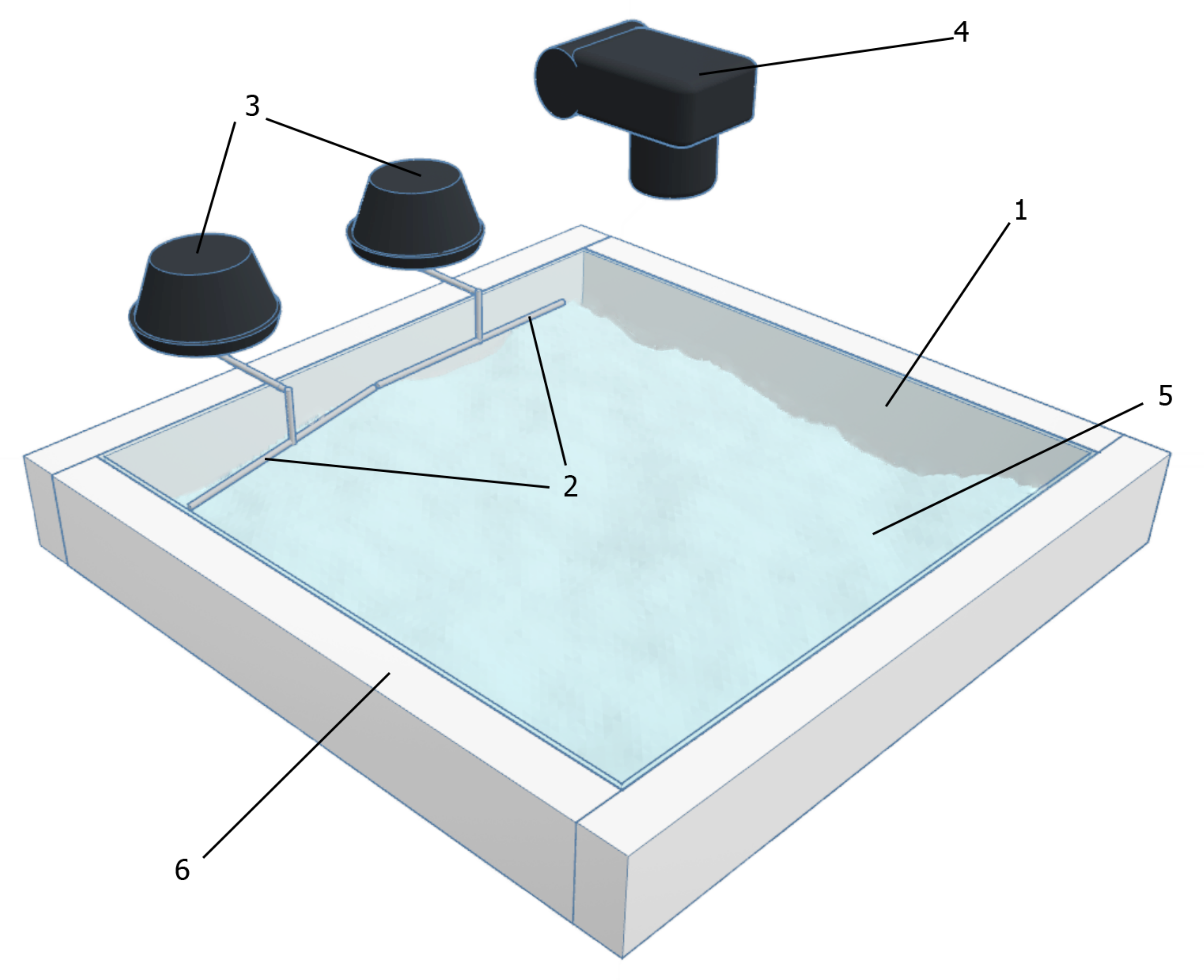}}
		\caption{Scheme of the experimental setup: (1) – bath, (2) – plungers, (3) – subwoofers, (4) – camera, (5) – waves on the surface, (6) – heat insulation.}
		\label{fig:1}
	\end{figure}

	The investigations were performed on experimental setup, the scheme of which is presented in Fig.~\ref{fig:1}. The setup consists of a bath with a length of $l=70$ cm, a width of $l=70$ cm, and a height of $25$ cm which was made from $10$-mm-thick glass. The bath was placed on a Standa vibration isolation table with an air suspension and was filled with distilled water to a depth of $d = 10$ cm. Two wave generators each consisting of an actuator and a plunger were mounted on a supporting frame; they excite waves on the water surface. The walls of the bath were isolated by styrofoam of thickness of $5$ cm to prevent the generation of thermally stimulated fluid flow.
	
	The plungers were made of a stainless steel rod with a diameter of $10$ mm and a length of $300$ mm. One of them was placed at an angle of $19^\circ$ to the bath wall. The second plunger was located parallel to the bath wall at a distance of $1$ cm from it. Two TS-W254R subwoofers (Pioneer) with a nominal power of $250$ W each were used as the plunger actuators. Sinusoidal signals with a frequency of $2.34$ Hz were generated by a dual-channel generator, amplified, and supplied to the subwoofers. The phase shift between the signals for subwoofers was equal to $\pi/2$. To visualize surface flow, a white powder of polyamide particles with an average diameter of about $30$ $\mu$m was deposited on the water surface. The particles on the surface were illuminated by the LEDs placed along the bath perimeter.

	Displacement of particles floating on the surface was captured by a Canon EOS $70$D camera at a rate of $24$ fps. The cross-correlation analysis of the image pairs using the PIVLab code \cite{thielicke2014pivlab} for MATLAB allowed us to obtain the horizontal velocity field associated with the tracers' motion and then calculate the vertical vorticity $\Omega$ of the slow current on the surface. Simultaneously, we recorded oscillations of the fluid surface in the vertical direction using a recently developed technique \cite{filatov2018technique}, which is based on restoring the surface curvature by analyzing the optical distortions of a contrast image at the bottom of the bath. The contrast image was illuminated by LEDs placed under the transparent bottom of the bath and was captured from the top by the same camera. The LED illumination under the bottom of the bath was synchronized with the even frames of video recording, while the LED illumination along the bath perimeter was synchronized with the odd frames.
	
	
	Applying the Fourier transform with the Hann window function to the obtained data, we find the distribution of vertical vorticity on the fluid surface and the wave spectrum in $k$-space. The Hann function falls off at the edges, so its application allows us to weaken the finite size effects and achieve periodic boundary conditions for the signals. Then we calculate the wave elevation and vorticity amplitudes based on the maximum values of corresponding peaks in $k$-space. The registration method and data processing algorithms were discussed in detail earlier, see Ref.~\cite{filatov2016generation}.

\section{Results}

\begin{figure*}
	\begin{minipage}[ht]{0.3\linewidth}
		\center{\includegraphics[width=\linewidth]{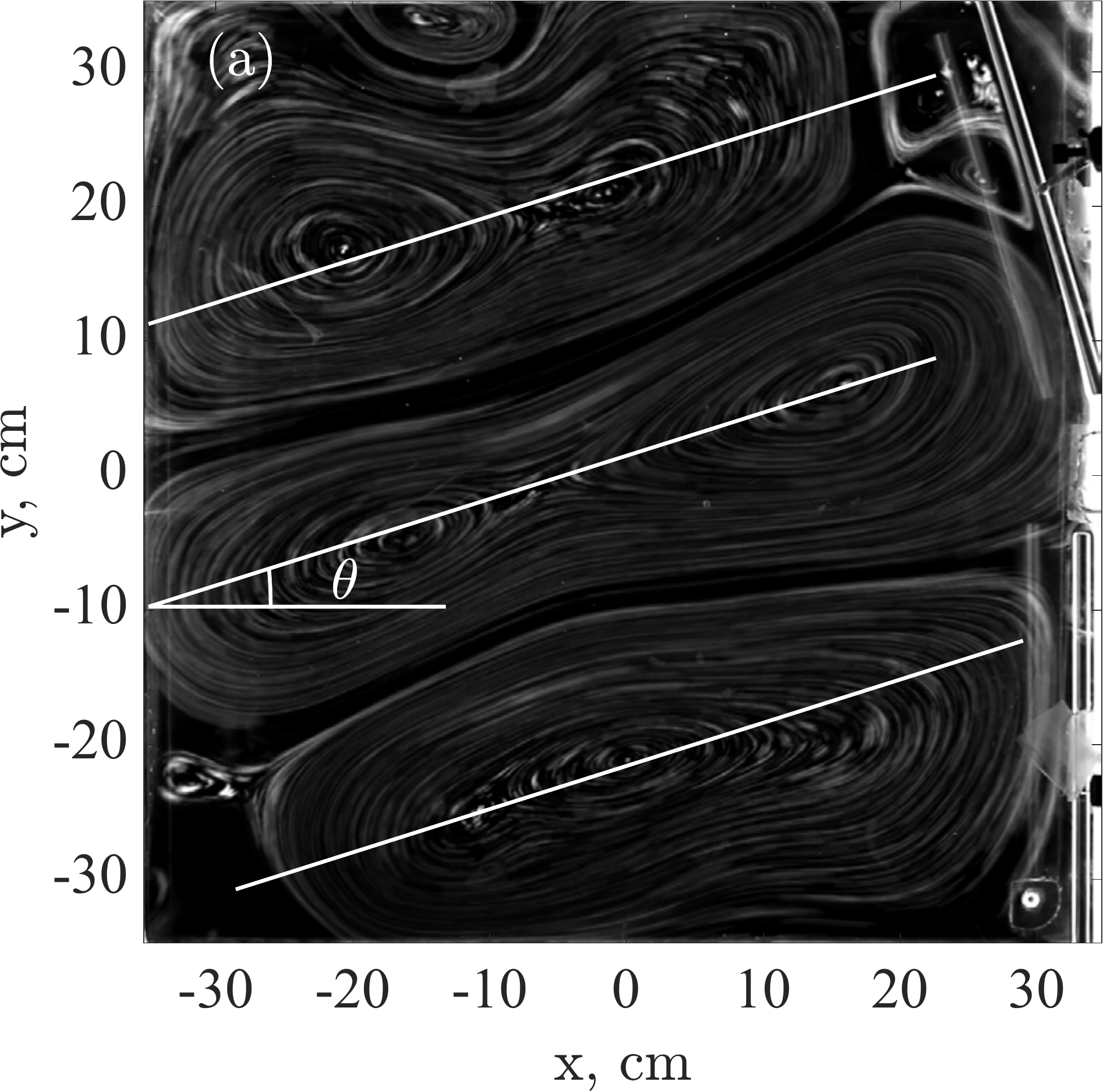}} 
	\end{minipage}
	\hfill
	\begin{minipage}[ht]{0.38\linewidth}
		\center{\includegraphics[width=\linewidth]{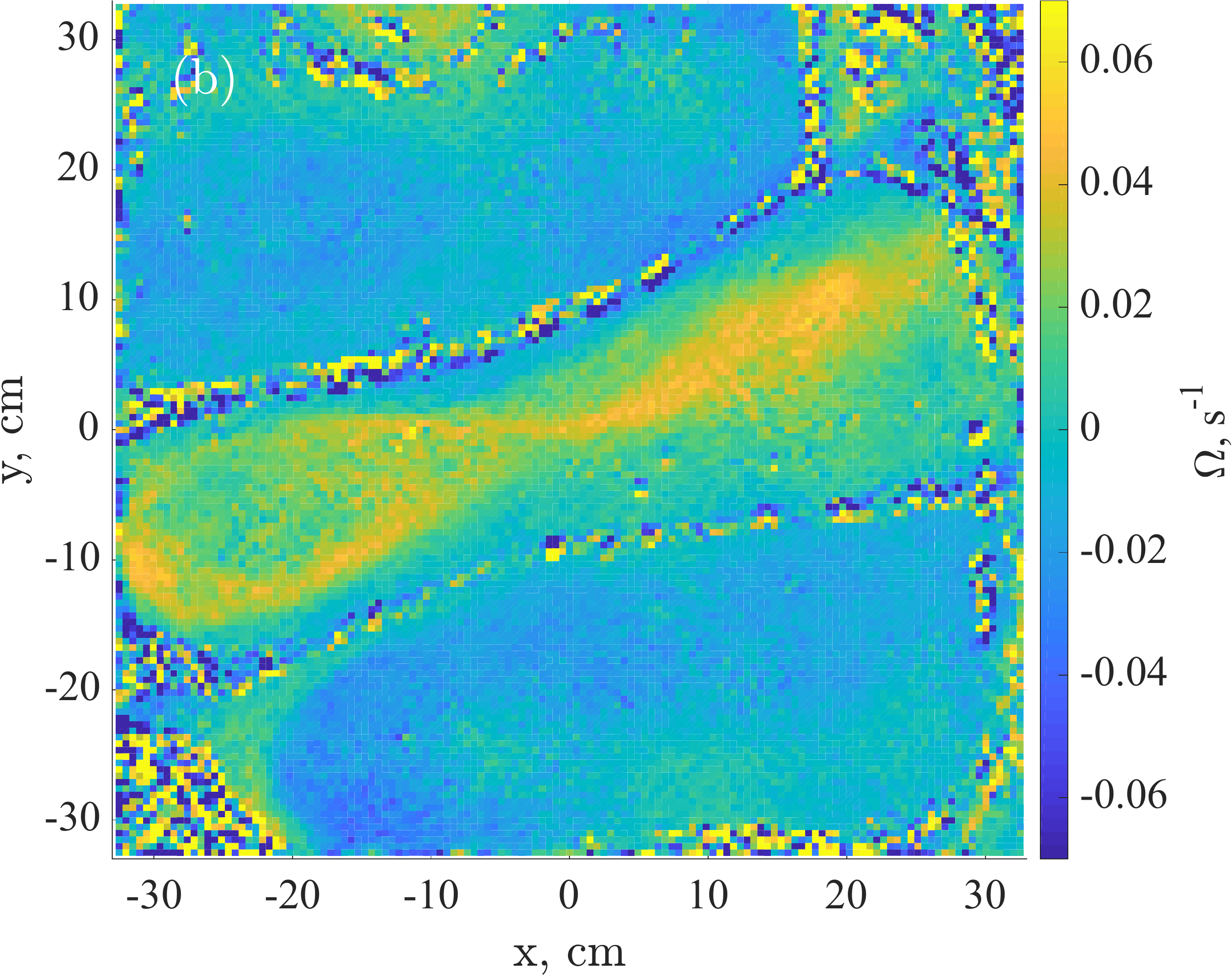}} 
	\end{minipage}
	\hfill
	\begin{minipage}[ht]{0.3\linewidth}
		\center{\includegraphics[width=\linewidth]{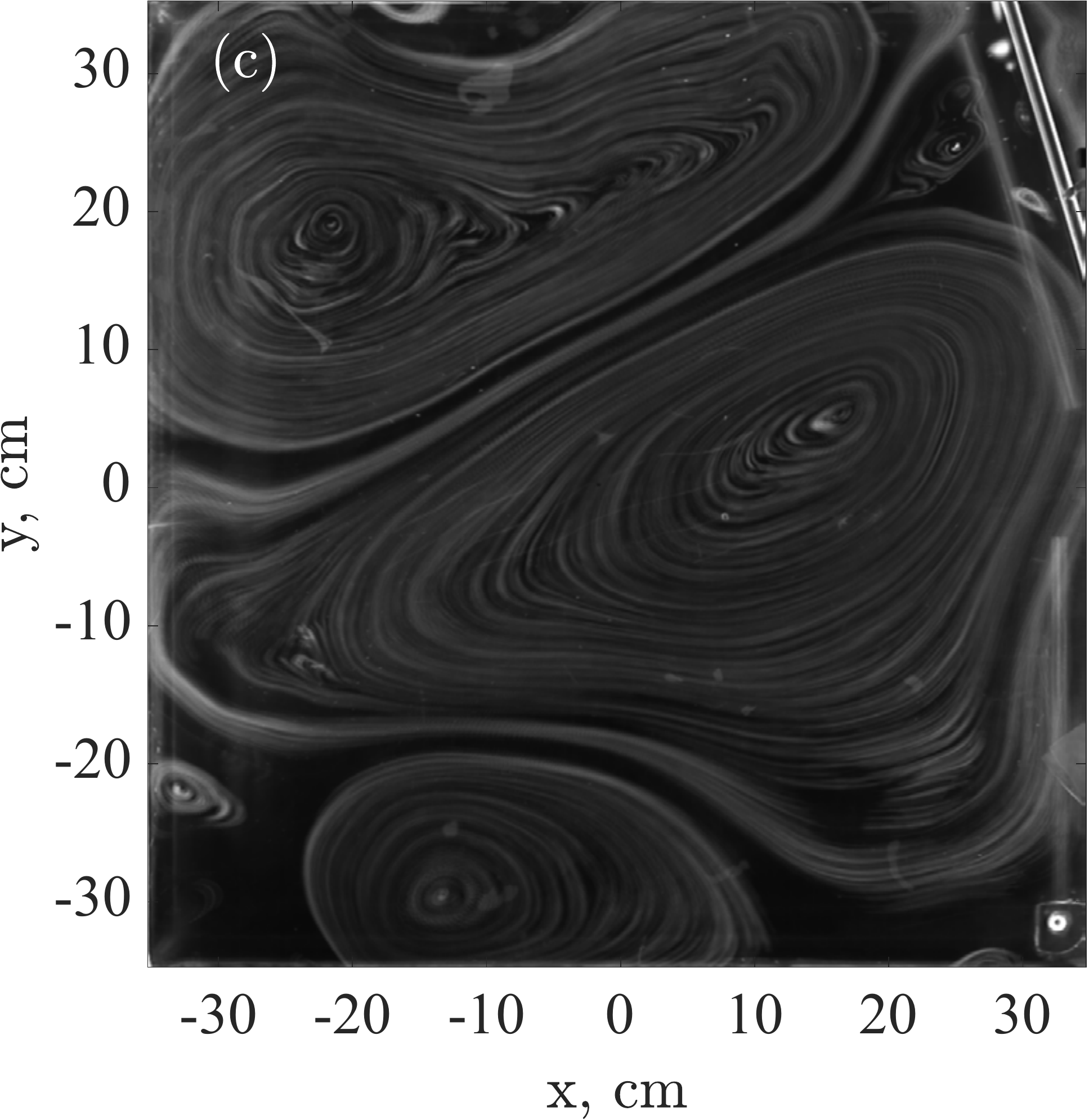}} 
	\end{minipage}

	\caption{(a) Tracks of polyamide particles floating on the fluid surface obtained $1000$ seconds after switching on the pumping. The amplitude of X-wave is $H_1 = 0.43$ mm and the effective Reynolds number $Re = \Omega_E t \approx 10$. (b) The corresponding distribution of vertical vorticity on the water surface. The measurements were averaged over three seconds. The mean vorticity is close to zero. (c) Tracks of polyamide particles at larger Reynolds number $Re = \Omega_E t \approx 20$ ($H_1 = 0.85$ mm and $t=1000$ seconds).}
	\label{fig:2}
\end{figure*}

At the initial moment, the fluid was at rest and we began to excite wave motion by applying sinusoidal signals of the same amplitude to both subwoofers. A few minutes later, three vortices become visible on the surface and their intensity increases over time. Fig.~\ref{fig:2}a shows the characteristic tracks of floating particles. The vortices are inclined at an angle of $\theta \approx 18^\circ$ to the $X$-axis and have an elongated shape. One can say that they form three stripes of width $L \approx 23$ cm each, and the length of the stripes is limited by the system size. Fig.~\ref{fig:2}b shows the corresponding distribution of vertical vorticity on the water surface. The middle strip has vorticity of the opposite sign in relation to the other two stripes. The pumping was turned off (the plungers stopped) $1054$ seconds after the start of the experiment. After that, the wave and vortex motion began to decay, and the video recording continued for another $512$ seconds. Fig.~\ref{fig:2}c shows the tracks of floating particles at the most intense level of pumping -- the spatial structure of vortices begins to distort due to the nonlinear interaction between them.
	
\begin{figure*}
	\begin{minipage}[ht]{0.49\linewidth}
		\center{\includegraphics[width=1\linewidth]{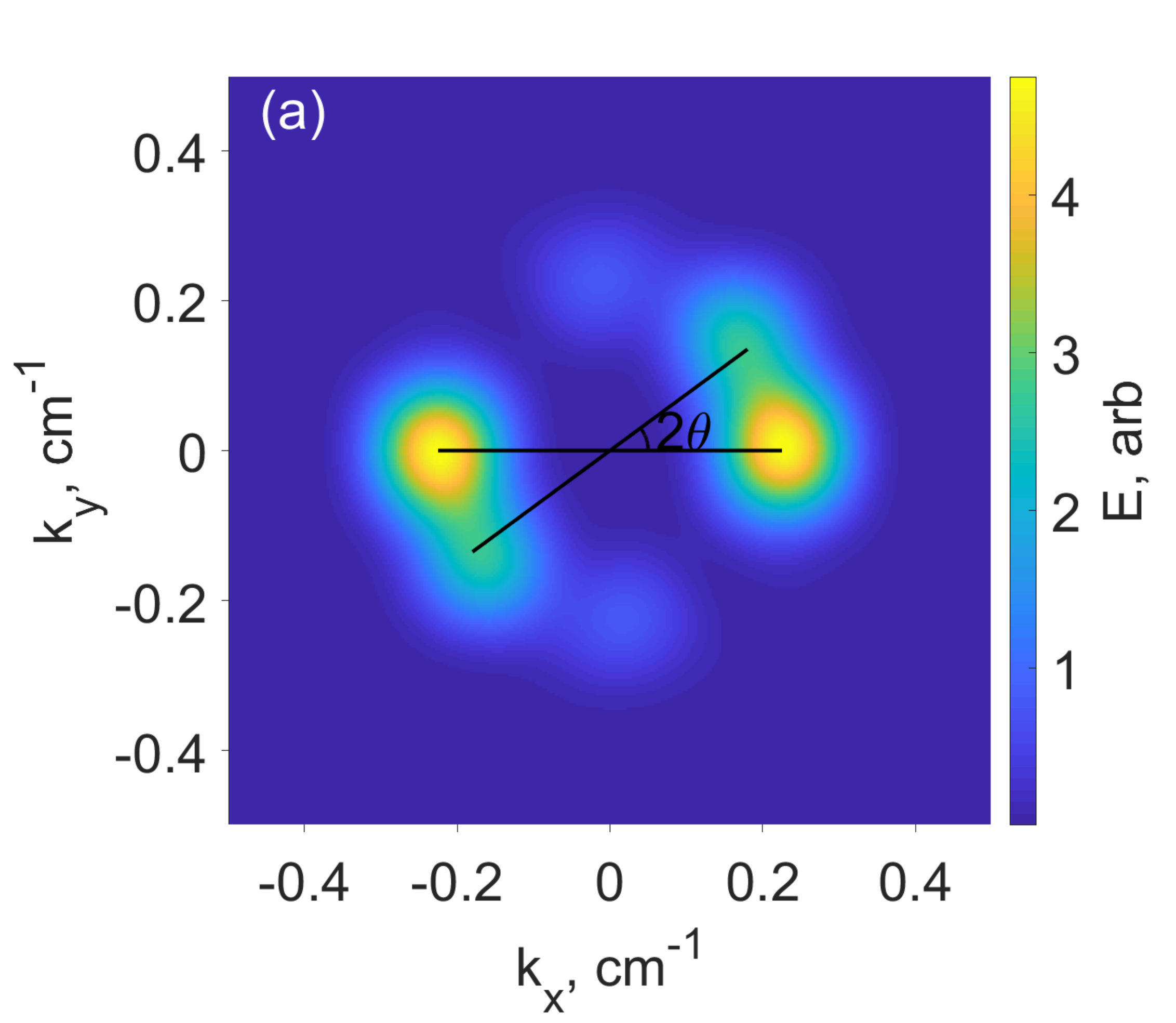}}
	\end{minipage}
	\hfill
	\begin{minipage}[ht]{0.49\linewidth}
		\center{\includegraphics[width=1\linewidth]{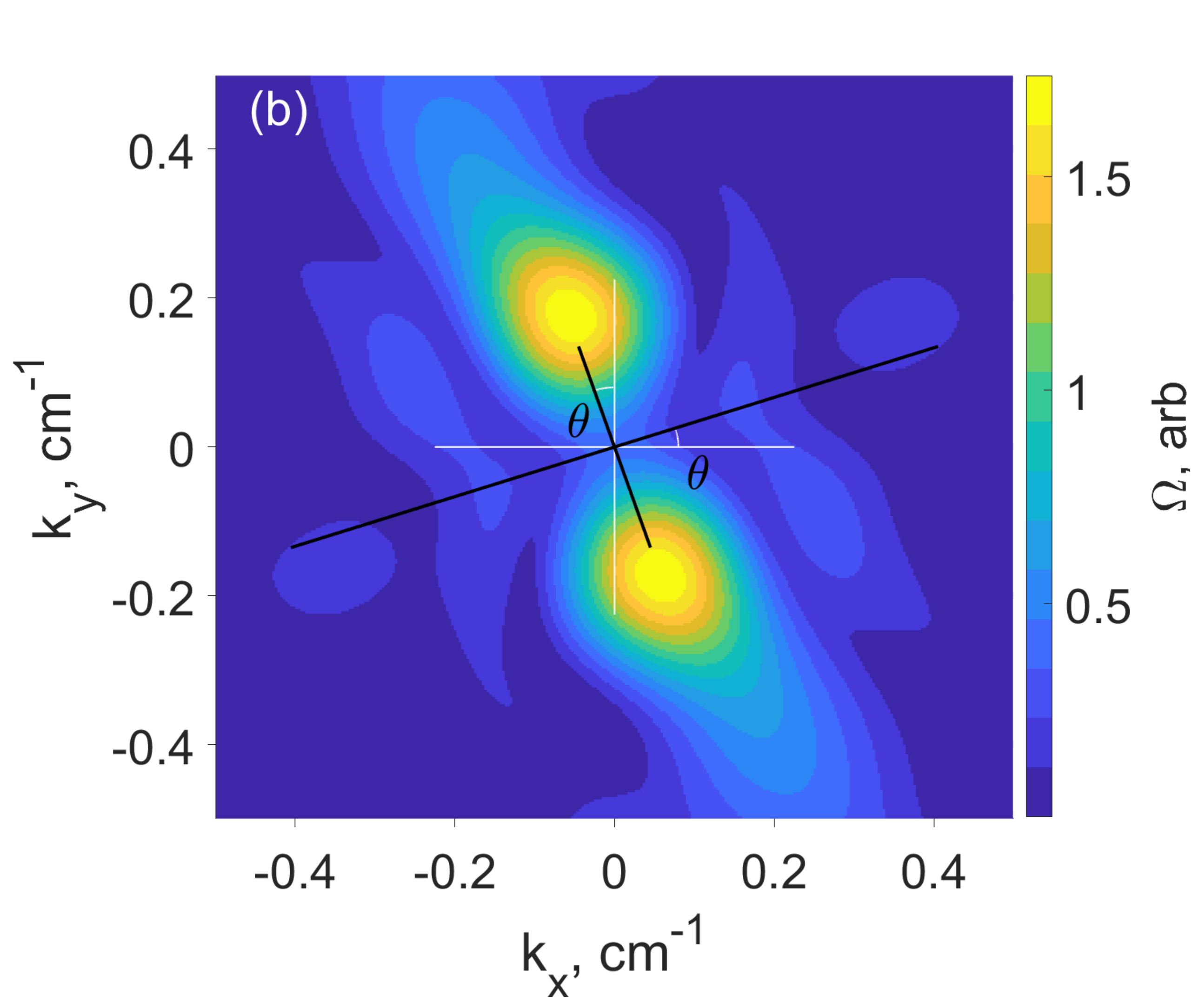}}
	\end{minipage}
	\caption{Distributions of (a) the wave energy and (b) the vertical vorticity in $k$-space obtained $1000$ seconds after switching on the pumping. The amplitude of X-wave is $H_1 = 0.43$ mm. The ends of the black lines in Fig.~\ref{fig:3}a mark the positions of the peaks calculated theoretically using the dispersion law $\omega^2 = gk \tanh(kd)$ and resonance conditions that determine the angle $2 \theta = \arctan(3/4)$. The ends of the black lines in Fig.~\ref{fig:3}b are located at distances $2k\sin \theta$ and $2k \cos \theta$ from the origin and correspond to the result of nonlinear interaction between excited waves.}
	\label{fig:3}
\end{figure*}

Fig.~\ref{fig:3}a demonstrates the energy distribution of surface vertical oscillations at a frequency of $2.34$ Hz in $k$-space, which corresponds to the time $t=1000$ seconds after the onset of wave excitation. These data describe the central region of the experimental cell as they were obtained using the Fourier transform with the Hann window function that allow us to reduce the influence of finite size effects. In the main approximation, one can assume that two standing waves are excited on the surface and their energies differ by several times. The wave number of both waves is equal to $k \approx 0.22$ cm$^{-1}$ that corresponds to the wavelength $\lambda \approx 28$ cm. Note that $kd \approx 2.2$ and therefore we may consider that the waves propagate on almost deep water \cite{landau1987course}.

The wave in the $X$-direction corresponds to the resonant mode of the system: $5$ half-waves fit over a bath length of $l=70$ cm, i.e. the wave vector $(k_x, k_y) = (5 \pi/l, 0)$. This wave is excited by a plunger that is parallel to the wall. Another plunger excites the wave with the wave vector $(k_x, k_y) = (4 \pi/l, 3 \pi/l)$. It spreads at an angle of $2 \theta = \arctan(3/4) \approx 36^\circ$ to the $X$-axis. Since the modes are degenerate in frequency, they are excited simultaneously. Let us emphasize that the direction of propagation of the second wave may not coincide with the perpendicular to the inclined plunger, since only resonant waves are excited in the system. Also, note that the angle of inclination of the vortices in Fig.~\ref{fig:2}a is half the angle between the excited waves.

Fig.~\ref{fig:3}b shows the distribution of the vertical vorticity in $k$-space. Two bright peaks correspond to an elongated vortex structure inclined at an angle of $\theta \approx 18^\circ$ to the $X$-axis and shown earlier in Fig.~\ref{fig:2}a. We also note the presence of two weaker peaks, which are indicated in the figure. The vertical vorticity is generated on the fluid surface as a result of the nonlinear interaction of two noncollinear waves discussed above, and therefore, in Fourier space, the vorticity distribution peaks are located at distances $2k\sin \theta$ and $2k \cos \theta$ from the origin. The width of the most intense vortex structure should be $L = \pi/(2k \sin \theta) \approx 23$ cm, which is in good agreement with the results presented in Fig.~\ref{fig:2}a.


\begin{figure*}
	\begin{minipage}[ht]{0.34\linewidth}
		\center{\includegraphics[width=\linewidth]{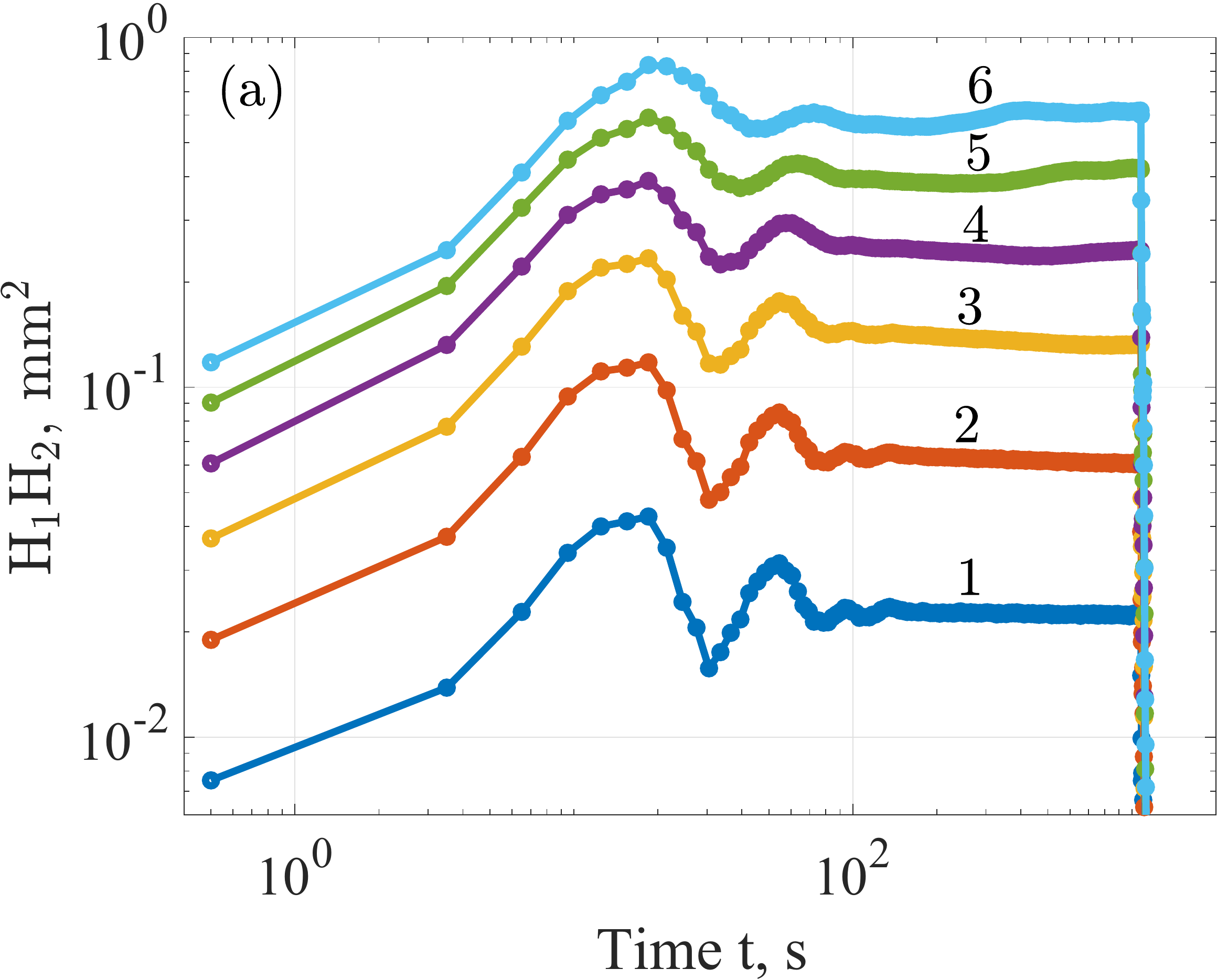}}
	\end{minipage}
	\hfill
	\begin{minipage}[ht]{0.34\linewidth}
		\center{\includegraphics[width=\linewidth]{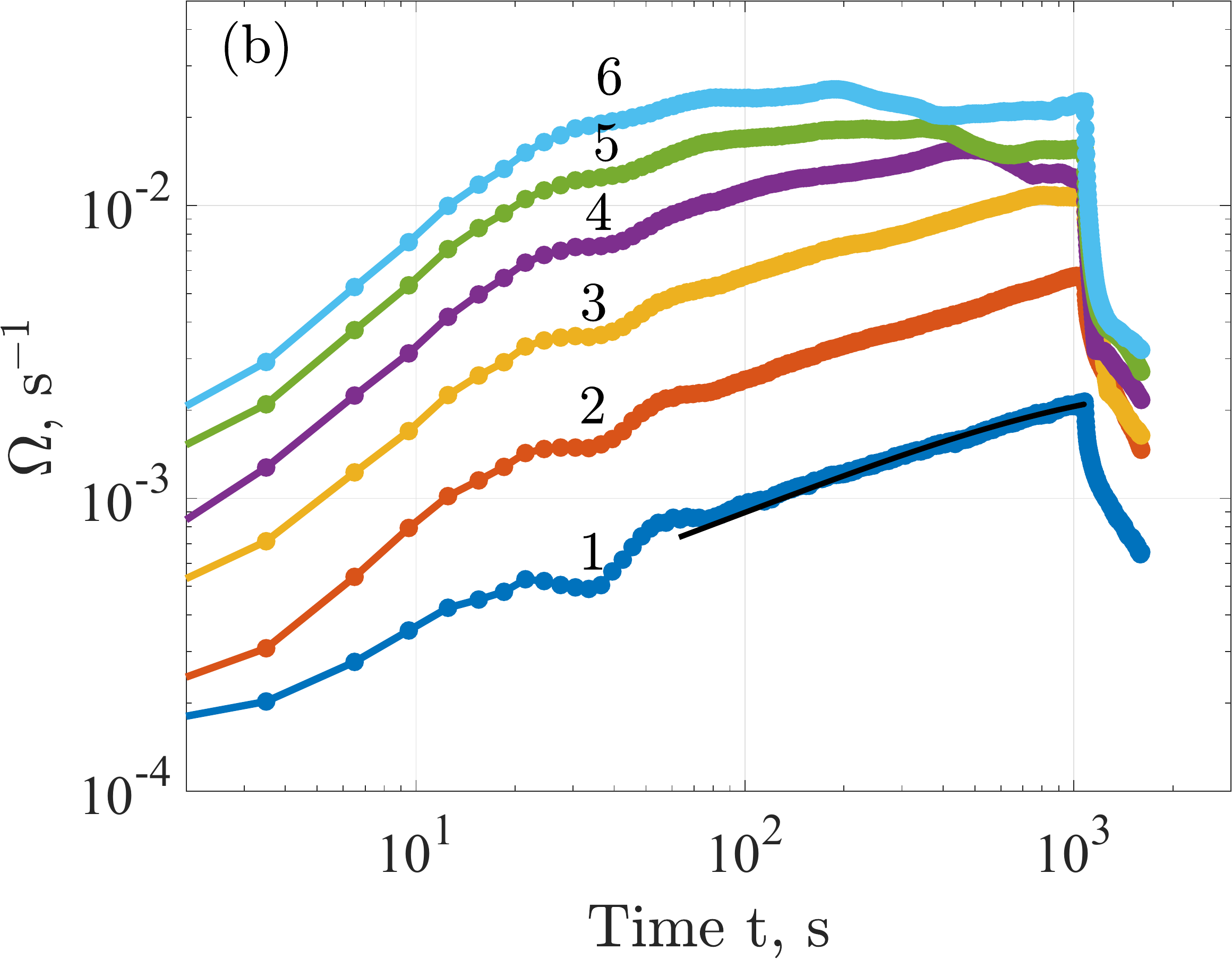}}
	\end{minipage}
	\hfill
	\begin{minipage}[ht]{0.3\linewidth}
		\center{\includegraphics[width=\linewidth]{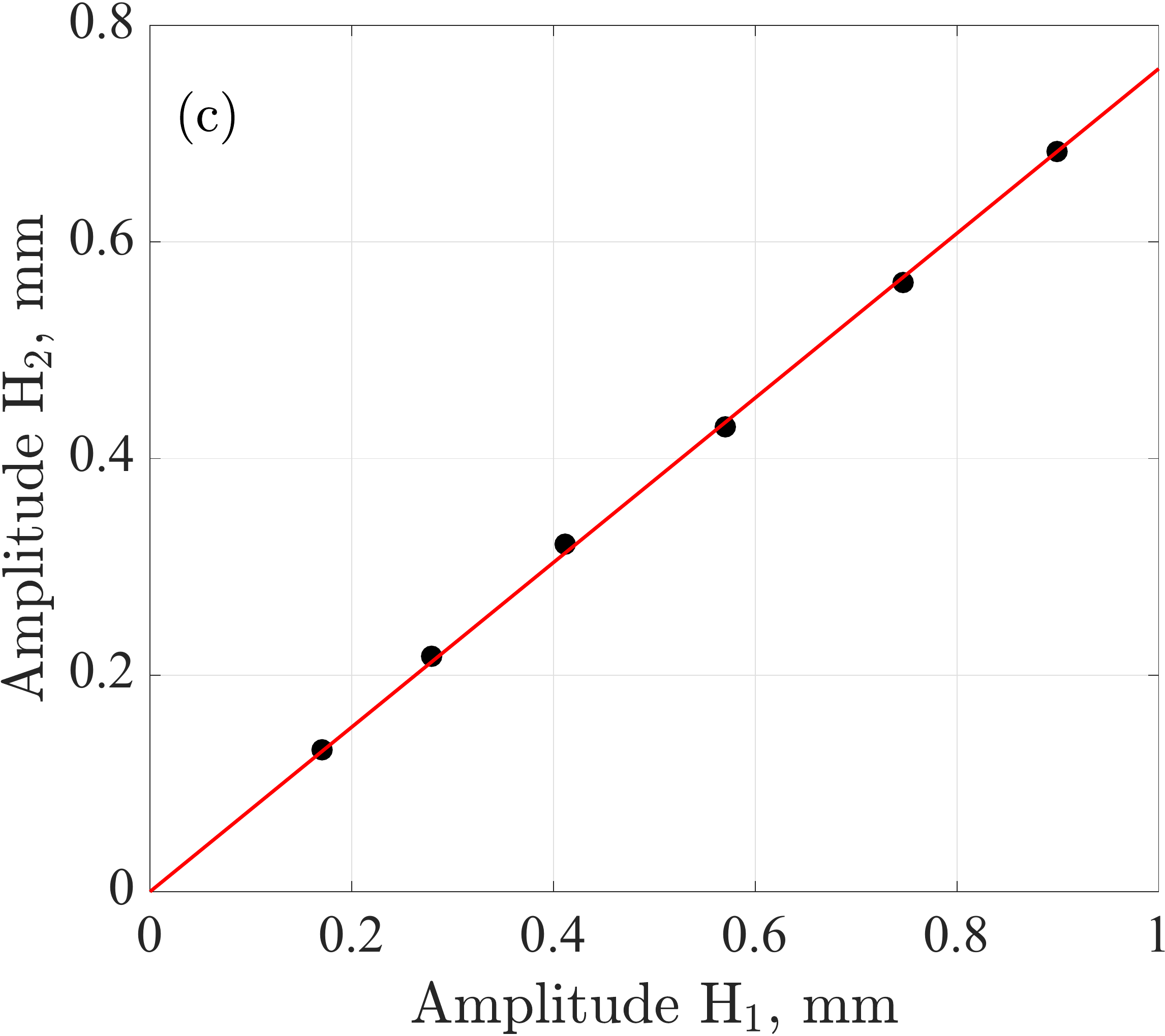}}
	\end{minipage}
	
	
	\caption{Time dependence of (a) the product of wave amplitudes $H_1 H_2 (t)$ and (b) the amplitude of vertical vorticity $\Omega(t)$ on the fluid surface. Different curves correspond to different values of $X$-wave amplitude $H_1$ in the stationary regime: 1 – 0.17 mm, 2 – 0.29 mm, 3 – 0.43 mm, 4 – 0.57 mm, 5 – 0.72 mm, 6 – 0.85 mm.  Fig.~\ref{fig:4}b also shows theoretical result corresponding to the model discussed in Sec.~\ref{sec:D} with the film compression modulus $\varepsilon = 0.37$ and $T_E=1001$ seconds (black line). Fig.~\ref{fig:4}c shows the dependence of the steady-state amplitude $H_2$ on $H_1$ at different pump levels. It turns out that $H_2 = 0.77 H_1$.}
	\label{fig:4}
\end{figure*}

Next, we calculate the wave elevation amplitudes $H_1$ and $H_2$, and the amplitude of vorticity $\Omega$ corresponding to the most intense vortex structure based on the maximum values of the peaks in $k$-space, and then investigate their dependence on time for different pumping intensities. Hereinafter we denoted $H_1$ -- the amplitude of the wave propagating in the $X$-direction, and $H_2$ -- the amplitude of the wave propagating at an angle of $2 \theta \approx 36^\circ$ to the $X$-direction.

Fig.~\ref{fig:4}a illustrates the dependence of $H_1 H_2$ on time for different pumping intensities. Standing waves are established within the first five seconds, a time that is approximately equal to the time of two-way propagation of waves from the plungers to the opposite walls and back. We recall that the group velocity of a wave with a frequency of $2.34$ Hz is approximately $35$ cm/s. The non-monotonicity of the dependence $H_1 H_2(t)$ is caused by a slight discrepancy between the pump frequency and the resonant mode frequency. The beats of the amplitude oscillations completely decay after about $100$ seconds, and the product of the wave amplitudes becomes constant.

Fig.~\ref{fig:4}b shows the dependence of $\Omega(t)$ starting from $2$ seconds after the start of the experiment. At times shorter than $100$ seconds it follows changes in the product $H_1 H_2 (t)$ of wave amplitudes. At longer times, the wave amplitudes do not change, but the vorticity $\Omega(t)$ continues to increase, i.e., we observe the production of the Eulerian contribution $\Omega_E(t)$ since the Stokes drift contribution $\Omega_S(t)$ remains at a constant level.

Fig.~\ref{fig:4}c demonstrates that the steady-state amplitude $H_2$ of the wave propagating at an angle of $2 \theta \approx 36^\circ$ to the $X$-direction changes synchronously with the steady-state amplitude $H_1$ when the pump level changes, and $H_2 = 0.77 H_1$.

\begin{figure}
	\centerline{\includegraphics[width=\linewidth]{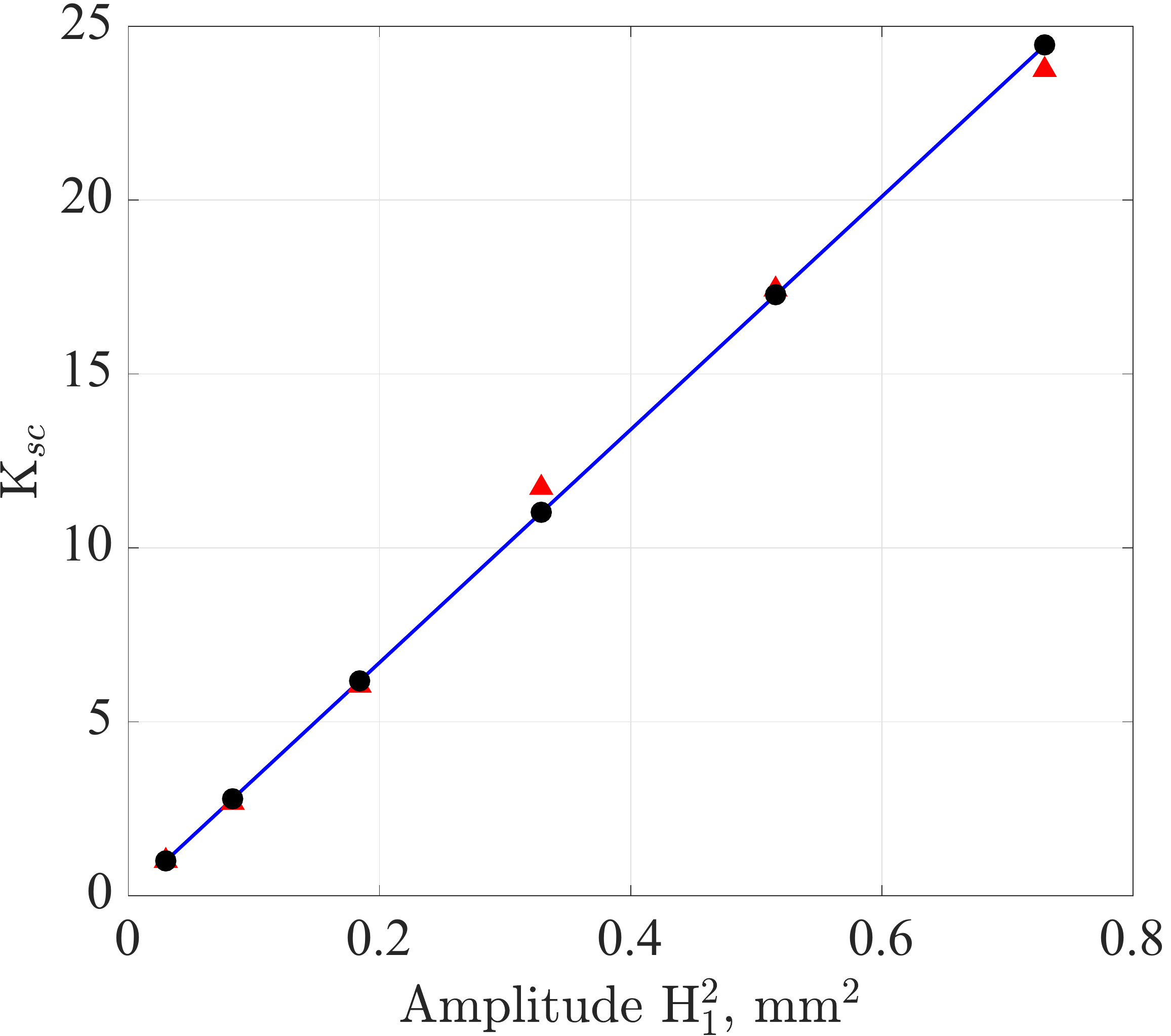}}
	\caption{Dependence of the scaling coefficient $K_{sc}$ on the steady-state amplitude $H_1$ at different pump levels. Marker $\Delta$ means that $K_{sc}$ was calculated using the dependence $H_1 H_2 (t)$, and $\bullet$ -– using the dependence $\Omega(t)$.}
	\label{fig:5}
\end{figure}

The curves shown in Figures ~\ref{fig:4}a and ~\ref{fig:4}b are similar, and it allows us to calculate the scaling factor, which will match all curves with the curve number $1$, that was obtained when the wave in the $X$-direction had a steady-state amplitude of $0.17$ mm. To determine the scaling factor we consider a time moment $\tau = 100$ seconds, when the wave amplitudes become stationary, and calculate the ratios
\begin{equation}
	K_{sc}^{(1)} =  \dfrac{H_1H_2(\tau, n)}{H_1 H_2(\tau, 1)}, \quad K_{sc}^{(2)} =  \dfrac{\Omega(\tau, n)}{\Omega(\tau, 1)},
\end{equation}
where $f(\tau, n)$ denotes the value $f(\tau)$ for the experimental curve with the number $n$. The results are presented in Fig.~\ref{fig:5}, which demonstrates that both ratios are well described by the same function, $K_{sc} = \beta H_1^2$, where the factor $\beta$ is equal to 33.5.

The rescaling results are presented in Fig.~\ref{fig:6}. The experimental points are shown starting from $3$ seconds, after the establishment of standing waves on the fluid surface. For relatively small wave amplitudes, the curves $\Omega(t)/K_{sc}$ coincide well with each other, and with an increase in the degree of nonlinearity, differences begin to appear. Good agreement of the rescaled curves at low steady-state wave amplitudes means that the generated vorticity $\Omega(t)$ is proportional to the product of steady-state wave amplitudes, since $K_{sc} \propto H_1^2$ and $H_2 \propto H_1$. A detailed discussion of the dependence $\Omega(t)$ in the case of weak nonlinearity is presented in the next section.

For large wave amplitudes, the degree of nonlinearity increases that leads to the interaction between vortices. The vortex shape is distorted, see Fig.~\ref{fig:2}c and the theoretical description stops working. Quantitatively, the degree of nonlinearity can be characterized by the effective Reynolds number $Re = \Omega_E(t) T_E$ for a slow current. The obtained results indicate that the higher-order nonlinearities becomes important at $Re \geq 10$. This is in line with previous measurements~\cite{parfenyev2019formation}. Note that nonlinear effects in standing waves are negligible in all experimental runs.

As for the decay stage, after the pumping is turned off, it can be seen from Figures \ref{fig:4}a and \ref{fig:4}b that the wave motion attenuates much faster compared to the vortex current. A detailed analysis of this stage makes it possible to determine the characteristic decay time of waves, which turns out to be equal to $\tau = 2 T_S \approx 30$ seconds. The vorticity relaxes much more slowly with a characteristic time close to $T_E \approx 1000$ seconds. This value is less than the theoretical estimate $T_E = L^2/(\pi^2 \nu) \approx 5 \times 10^{3}$ seconds, which apparently can be explained by the friction of vortices against the boundaries of the system.
	
	\section{Discussion}\label{sec:D}
	
	The formation of vortex flow by noncollinear waves on the surface of deep water was studied theoretically in Ref.~\cite{parfenyev2020large}. Let us direct the $X'$-axis along the direction of the stripes shown in Fig.~\ref{fig:2}a and the $Y'$-axis -- perpendicular to it. In this coordinate system, the surface elevation can be written in the form
	\begin{equation}
		\begin{aligned}
			h(t,x',y') = H_1 \cos(\omega t) \cos(kx' \cos \theta - ky' \sin \theta) + \\
			+ H_2 \cos (\omega t + \psi) \cos(kx' \cos \theta + ky' \sin \theta),
		\end{aligned}
	\end{equation}
	where $\psi$ is the phase shift between excited waves, which can be controlled by changing the phase shift between the electrical signals applied to the plunger drives. In our case, the parameters are chosen so that the value of $\psi$ is close to $-\pi/2$.
	
	In what follows, we are interested in the vertical vorticity, which corresponds to the two brightest peaks in Fig.~\ref{fig:3}b. The contribution $\Omega_S$ associated with the Stokes drift is equal to
	\begin{equation}\label{eq:S}
		\Omega_S = \omega k^2 H_1 H_2 \sin 2 \theta \cos^2 \theta \cos [2ky' \sin \theta],
	\end{equation}
	see \cite[Eq.~(E2)]{parfenyev2020large}. 
	As for the Eulerian contribution to the vertical vorticity, since the wave motion is established quickly compared to the vortex flow, $T_S \ll T_E$, and since the duration of the experiment is short compared to the characteristic time of viscous diffusion to the bottom of the system, $t \ll d^2/\nu$, the following expression is valid
	\begin{equation}\label{eq:E}
		\begin{aligned}
			\Omega_E = \left(2 + \dfrac{\varepsilon^2/\cos^2 \theta}{2 \sqrt{2} \gamma (\varepsilon^2 - \varepsilon \sqrt{2} + 1)} \right) \omega k^2 H_1 H_2\times \\ \times\cos^3 \theta
			\cos [2ky' \sin \theta] \, \mathrm{erf} \left[ \sqrt{t/T_E}\right].
		\end{aligned}
	\end{equation}
	Here we introduced $\gamma = \sqrt{\nu k^2/\omega}$ and assumed that the fluid surface might be covered with a thin insoluble liquid film due to contamination \cite{campagne2018impact}. The presence of such a film on the fluid surface can significantly increase the intensity of vortex flows, which was demonstrated in Ref.~\cite{parfenyev2019formation} for orthogonal standing waves. In the simplest model, the properties of the film can be described by only one parameter -- the dimensionless compression modulus $\varepsilon \geq 0$ introduced in Ref.~\cite{parfenyev2018influence} and characterizing the film properties at frequency $\omega$ and scale $1/k$. The limiting case of a free surface corresponds to $\varepsilon \rightarrow 0$ and in the opposite case $\varepsilon \rightarrow \infty$ we deal with an almost incompressible surface film.

	\begin{figure}
		\centerline{\includegraphics[width=\linewidth]{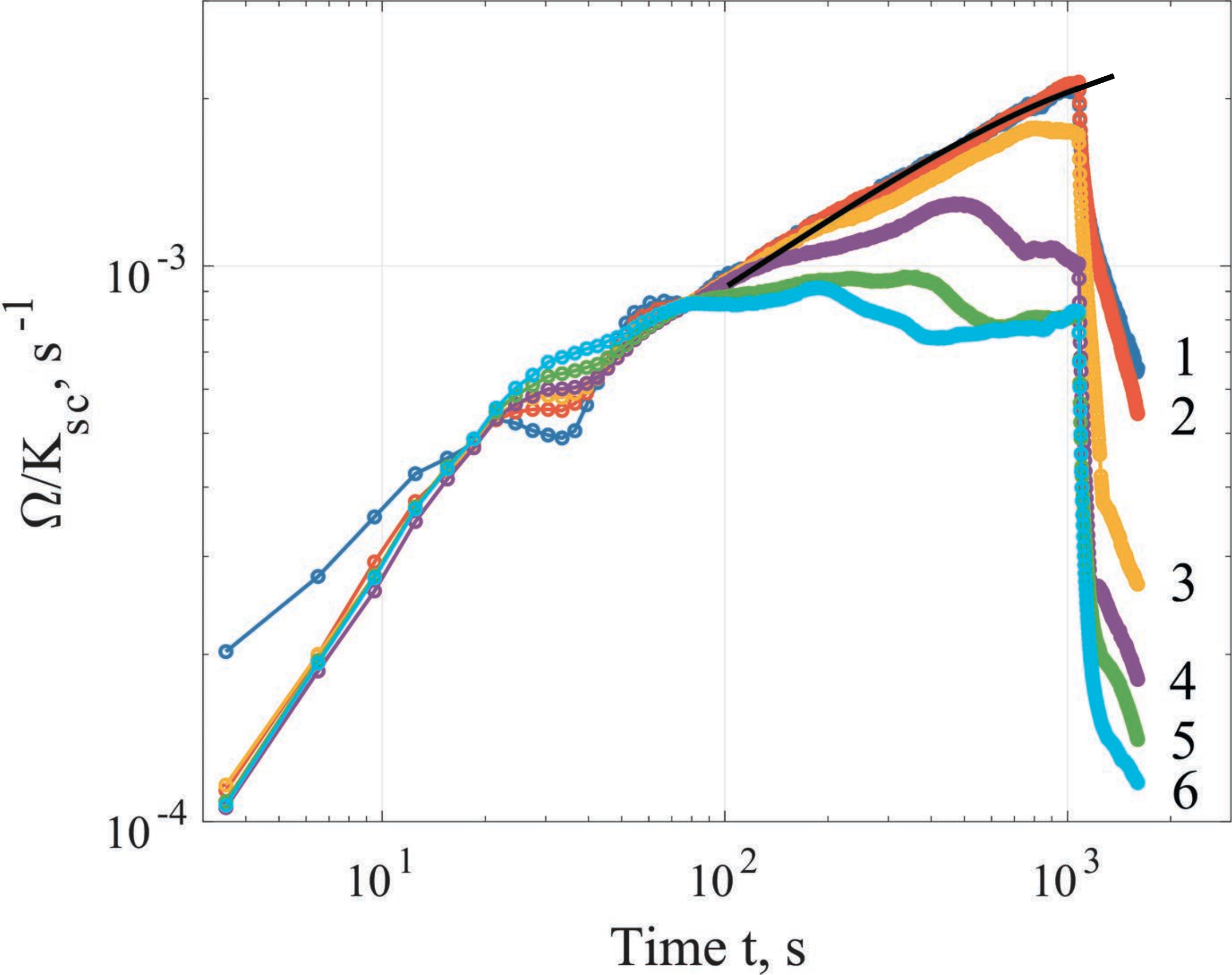}}
		\caption{Time dependence of the rescaled vorticity $\Omega(t)/K_{sc}$ at different pump levels expressed in terms of the steady-state $X$-wave amplitude $H_1$: 1 – 0.17 mm, 2 – 0.29 mm, 3 – 0.43 mm, 4 – 0.57 mm, 5 – 0.72 mm, 6 – 0.85 mm. The solid black line corresponds to the theoretical prediction, see Sec.~\ref{sec:D}, with the compression modulus of the film $\varepsilon = 0.37$ and $T_E=1001$ seconds.}
		\label{fig:6}
	\end{figure}
	
	Next, we use equation (\ref{eq:1}) together with relations (\ref{eq:S}) and (\ref{eq:E}) to explain the measured dependence $\Omega(t)$. The results are shown in Fig.~\ref{fig:4}b and in Fig.~\ref{fig:6} with solid black curves. The best fit corresponds to the film compression modulus $\varepsilon = 0.37$ and $T_E=1001$ seconds, which is in agreement with the results obtained from the analysis of the decay stage. We also note that the presence of a film on the liquid surface enhances the dissipation rate of the wave motion after switching off the pumping. To estimate the expected time of wave decay $\tau$, we shall use the relation~\cite[Eq.~(5)]{parfenyev2019formation}
	\begin{equation}\label{eq:dissipation}
		\frac{1}{\omega \tau}
		=  2\gamma^2 \left(1 + \dfrac{1}{\gamma kl \sqrt{2}} \right) +\frac{\gamma}{2\sqrt{2}}\frac{\varepsilon^2}{(\varepsilon^2-\varepsilon\sqrt{2}+1)},
	\end{equation}
	where $l=70$ cm is the length of the side of the square experimental cell and the corresponding term takes into account the dissipation near the system boundaries \cite[\S25]{landau1987course}. After straightforward calculations, we find $\tau \approx 63$ seconds. This value is larger than the experimental result, but they agree in order of magnitude. We believe that this difference is caused by some other energy dissipation mechanisms, which were not taken into account theoretically (e.g., the friction of fluid against the plunger which remained partially submerged in the fluid after the pumping was turned off). As regards the value of the film compression modulus $\varepsilon = 0.37$ , it is in good agreement with the results reported earlier \cite{parfenyev2019formation}.

Finally, let us estimate the relative contributions of the Stokes drift and the Eulerian term to the vertical vorticity. At $100$ seconds and the wave amplitude $H_1 = 0.43$ mm the Eulerian contribution $\Omega_E \approx 5 \times 10^{-3} \, s^{-1}$, while the Stokes drift $\Omega_S \approx 5 \times 10^{-4} \, s^{-1}$, i.e. $\Omega_E/\Omega_S \approx 10$. At $1000$ seconds and at the same wave amplitude, the Euler contribution is $\Omega_E \approx 10^{-2} \, s^{-1}$, while the Stokes drift remains the same, and it means that now $\Omega_E/\Omega_S \approx 20$. Note that the ratio of the Euler and Stokes contributions to the vertical vorticity does not depend on the wave amplitudes if the effective Reynolds number for the slow current is small, since both terms are proportional to the product $H_1 H_2$, see Eqs.~(\ref{eq:S}) and (\ref{eq:E}).

	\section{Conclusion}
	
	It was shown experimentally that noncollinear waves propagating at an angle $2 \theta$ on the water surface generate a stripe-like vortex flow. The width of the stripes can be estimated as $L=\pi/(2k \sin \theta)$ and their length is limited by the system size. The vertical vorticity $\Omega$ of the flow is proportional to the product of wave amplitudes and can be represented as a sum of Stokes drift and Eulerian contributions, $\Omega = \Omega_S + \Omega_E$, see expressions (\ref{eq:S}) and (\ref{eq:E}).
	
	The Stokes drift instantly tracks changes in wave amplitudes, and therefore the rate of change of this contribution with time is determined by the rate of establishment of wave motion. At large times, the contribution $\Omega_S$ to the total vertical vorticity $\Omega(t)$ turns out to be small.
	
	The Eulerian contribution, on the contrary, has a relatively slow kinetics and its growth with time is well described by the dependence $\Omega_E(t) \propto \mathrm{erf} \left[ \sqrt{t/T_E}\right]$. This contribution leads to an increase in the total vorticity $\Omega(t)$, which can be seen in Fig.~\ref{fig:4}b at times exceeding the establishment time of the wave motion. We also found that the experimentally measured time $T_E$ is much less than the theoretical estimate $T_E = L^2/(\pi^2 \nu)$, which is apparently due to the friction of the vortices against the side walls of the system.
	
	A quantitative agreement with the theory was achieved only by taking into account possible contamination of the fluid surface \cite{parfenyev2018influence}. In this case, the amplitude of the Euler contribution at the surface increases significantly, see expression (\ref{eq:E}). The found value for the parameter $\varepsilon = 0.37$ that describes the elastic properties of the contaminated surface is in close agreement with the results reported earlier in Ref.~\cite{parfenyev2019formation}. Overall, the presented experimental results are well described by the existing theoretical model of the phenomenon.
	
	\section{Acknowledgements}
	
	The work was supported by the Russian Ministry of Science and Higher Education, project No. 075-15-2019-1893.

\end{document}